# Micro-economic Analysis of the Physical Constrained Markets: Game Theory Application to Competitive Electricity Markets

Ettore Bompard , Yuchao Ma , Elena Ragazzi

*Abstract*- **Competition has been introduced in the electricity markets with the goal of reducing prices and improving efficiency. The basic idea which stays behind this choice is that, in competitive markets, a greater quantity of the good is exchanged at a lower and a lower price, leading to higher market efficiency.**

**Electricity markets are pretty different from other commodities mainly due to the physical constraints related to the network structure that may impact the market performance. The network structure of the system on which the economic transactions need to be undertaken poses strict physical and operational constraints.**

**Strategic interactions among producers that game the market with the objective of maximizing their producer surplus must be taken into account when modeling competitive electricity markets. The physical constraints, specific of the electricity markets, provide additional opportunity of gaming to the market players. Game theory provides a tool to model such a context. This paper discussed the application of game theory to physical constrained electricity markets with the goal of providing tools for assessing the market performance and pinpointing the critical network constraints that may impact the market efficiency. The basic models of game theory specifically designed to represent the electricity markets will be presented. IEEE30 bus test system of the constrained electricity market will be discussed to show the network impacts on the market performances in presence of strategic bidding behavior of the producers**.

## I. INTRODUCTION

Nowadays the liberalization of the power industry has been implemented in many countries. The introduction of the deregulation has not always proved to be as efficient as expected. In California [1, 2], the market experienced huge problems. From May 2000 to May 2001, the price hit frequently the cap and forced the regulator to revise the price cap downward. The average price of December 2000 was 317 $/MWh, almost ten times higher than usual. In June 1998, wholesale electricity price in Midwest US market reached 7,000 $/MWh [3]. Starting from the regulated monopoly, the competition in the electricity markets was aimed to improve market efficiency toward the theoretical reference model of perfect competition. Actually, due to the structural characteristics, the electricity markets are oligopoly in which the market performances are in-between perfect competition and monopoly. In this context, the task of the regulators is to force them toward perfect competition while monitoring continuously the distance from such a condition, or to avoid market power exploitation.

In the electricity markets, as well as in other markets, market power may arise striving for larger amount of profits or surpluses with high prices and capacity withdrawals, compared with the competitive values [4]. Game theory [5-6] can capture the strategic interactions among producers who are aware that their results depend on other competitors' decisions. Based on the game theory, [10-22] investigated the strategic interactions among producers in electricity markets.

In addition to the traditional causes of market power, in the electricity markets, the network constraints may give additional possibilities of market power behaviors arising that are very specific of this contest. An instantaneous balance between power injected by the generators and the power withdrawn by loads plus the transmission losses should be guaranteed to keep the system frequency at the rated value. The Kirchhoff laws must be satisfied and a power balance at each bus must be enforced. The power flow paths, directions and values, are depended on the bus voltage profile and primary constants of the transmission lines and those lines have flow limits including thermal, voltage drop and stability limits. In addition, from the operational point of view, the voltage profile of the system must be kept within a specific interval. Therefore, the power systems that accommodate the economic transactions in the market need to be operated under strict physical and operational constraints to assure its feasibility; if these constraints are binding the system is said to be congested and proper measures need to be undertaken [7]. This paper is aimed to discuses the network constraints impacts on the market performances under oligopoly models.

This paper consists of four additional sections. In section II, the market clearing model under network constraints is introduced. Section III discusses different game models while in section IV the numerical studies with respect to IEEE30 bus system is presented .Section V provides some conclusive remarks.

This research has been supported by the European Commission under grant ASI/B7-301/98/679-026-ECLEE project and HERMES - Higher Education and Research on Mobility regulation and Economics of local Services – Turin.

Ettore Bompard, Yuchao Ma are with Politecnico di Torino, Department of Electrical Engineering - Italy , ettore.bompard@polito.it
Elena Ragazzi is with CERIS, Institute for Economic Research on Firms and Growth –CNR, National Research Council, e.ragazzi@ceris.cnr.it

## II MARKET CLEARING MODEL

In the pool operated electricity markets, the Independent System Operator (ISO) takes the responsibility of coordinating the aggregate offers from the supply side and the aggregate demand curves for a specified time interval trading, usually one hour. That leads to the determination of market equilibrium, characterized by a unique market clearing price ($\lambda$) and a market clearing quantity ($q$) (Fig.1 left). The social surplus is composed by the consumer surplus ($S^C$) and producer surplus ($S^G$). However, due to the peculiarities of the electricity transmission, the transactions must be settled according to the physical constraints of the electricity network and different nodal prices may arise when the flow limits are binding ($\lambda''$ and $\lambda'$ are respectively for the demand side and supply side, Fig.1 right). In this case, merchandise surplus ($S^M$) will arise, the area $\lambda''EA\lambda'$ (Fig.1 right). The social surplus is equal to the summation of the consumer surplus, merchandise surplus and producer surplus.

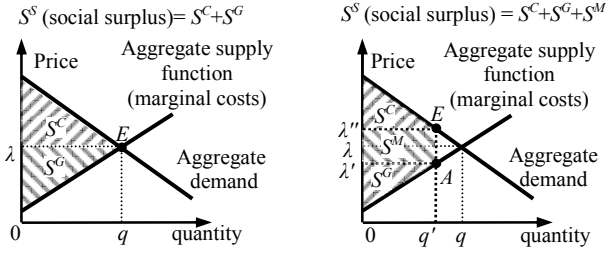

Fig.1. the market clearing without (left) and with (right) network constraints

A producer that is unable to exercise market power is known as price taker. According to the classic economic theory, a price-taking producer that wishes to maximize his profits would bid his power production at his own marginal cost and the market is characterized as perfect competition [8-9].

In perfect competition markets a large number of price-taking producers with a very small market share produce homogeneous and perfectly substitutable commodities. Furthermore the market should not have significant entry barriers but have free mobility of all the related production resources and perfect information among producers. Although the perfect competition is completely unrealistic, it can serve as a reference case to identify market power behaviors in a practical market, basing on the fact of that perfect competition would lead to the most efficient market performance.

Network constraints distinguish the electricity markets from most of other commodity markets. Considering the network constraints, the perfect competition market equilibrium can be interpreted as an optimization problem. Let's assume that at each generator bus there is just one producer (generator), the cost function of the producer $g$ is:

$$C_g(p_g) = a_g^m p_g + \tfrac{1}{2} b_g^m p_g^2 \qquad \forall g \in G \qquad (1)$$

and the marginal cost of the producer $g$ is:

$$c_g(p_g) = a_g^m + b_g^m p_g \qquad \forall g \in G \qquad (2)$$

where $a_g^m$ and the $b_g^m$ are respectively the intercept(\$/MW) and slope(\$/MW$^2$) of the marginal cost function; $p_g$ is the production quantity (MW); $G$ is the set of generator buses.

At the load bus $d$, the electricity consumer $d$ is modeled with a linear demand function:

$$v_d = e_d + h_d q_d \qquad \forall d \in D \qquad (3)$$

where $e_d$ and the $h_d$ are respectively the intercept (\$/MW) and slope (negative, \$/MW$^2$) of the demand function; $q_d$ is the demand quantity of the load $d$ (MW); $D$ is the set of load buses.

The market clearing based on the linear DC power flow can be formulated as:

$$\max \quad S^S = \tfrac{1}{2} q^T H q + q^T e - (\tfrac{1}{2} p^T B^m p + p^T a^m) \qquad (4)$$

$$s.t. \qquad I_G^T p - I_D^T q = 0 \qquad \leftrightarrow v_N \qquad (5)$$

$$-T \leq J(p-q) \leq T \qquad \leftrightarrow \mu^+/\mu^- \qquad (6)$$

$$P^{min} \leq p \leq P^{max} \qquad \leftrightarrow \omega^+/\omega^- \qquad (7)$$

where:
- $p$ : power production vector
- $q$ : power demand vector
- $e, h$: intercept and slope parameter vectors for linear demand curves
- $a^m, b^m$: intercept and slope parameter vectors for linear marginal costs
- $H, B^m$: diagonal matrix (diagonal elements: the vector $h$ or $b^m$, respectively)
- $P^{min}, P^{max}$: vectors of lower, upper capacity (MW) for the generators
- $J$: power transfer distribution matrix
- $T$: the flow limits (MW) vector
- $I_G, I_D$: identity vector (same dimension as the power or demand vector

The superscript "$T$" is used to denote the transpose operation for the matrices or vectors.

The equality expression (5) is for the power balance, the associated Lagrange multiplier $v_N$ is the nodal price at the reference bus $N$. The inequality expressions (6) and (7) represent the line flow limits and the power generation lower and upper limits; $\mu^+/\mu^-$ and $\omega^+/\omega^-$ are the associated Lagrange multiplier vectors for the line flow limits and for the generation limits.

The nodal prices ($\lambda$) at the buses other than the reference bus $N$ can be expressed as linear functions in terms of the $v_N$ and $\mu^+/\mu^-$:

$$\lambda = f(v_N(p,q), \mu^+, \mu^-) \qquad (8)$$

When the line flow are not binding, $\mu^+ = \mu^- = 0$

$$\lambda = v_N(p,q) \qquad (9)$$

all the nodal prices are equal to the reference bus price.

The power production $p$ and the load demanded $q$ are:

$$p = (B^m)^{-1}[\lambda_G - (\omega^+ - \omega^-) - a^m] \qquad (10)$$

$$q = H^{-1}(\lambda_D - e) \qquad (11)$$

When the production are not binding at the capacity limits ($\omega^+ = \omega^- = 0$), the nodal prices of the generators are at their supply curves (marginal cost curves under perfect competition). As for the loads, the nodal prices are at the demand curves. Provided with the nodal prices and the power quantities, the producer surplus ($S_g^G$) and the consumer surplus ($S_g^C$) can be expressed as:

$$S_g^G = \lambda_g p_g - (a_g^m p_g + \tfrac{1}{2} b_g^m p_g^2) \quad \forall g \in \mathcal{G} \tag{12}$$

$$S_d^C = e_d q_d + \tfrac{1}{2} h_d q_d^2 - \lambda_d q_d \quad \forall d \in \mathcal{D} \tag{13}$$

However, the perfect competition is just an ideal market that serves as reference case. The electricity market is closer to the oligopoly model in which the producers may exert market power behaviours in presence of strategic biddings to maximize their producer surpluses. Modelling the oligopoly market clearing is done by substituting the marginal cost curves in the objective function of the perfect competition model (4) with the strategic biddings of the producers and the object function value is called system surplus.

## III. OLIGOPOLY COMPETITION MODELS: GAME THEORY APPLICATIONS

Game theory was founded in 1944 by Von Neumann and Morgenstern. The papers written by Nash in 1951 on the definition and existence of Equilibrium are the basis for modern non cooperative game theory. In the last 50 years game theory has become a crucial tool for the analysis of strategic behaviors of individuals and competition among companies in oligopoly markets. For electricity markets, basic game "ingredients" are:

*Game*: is a set of rules that discipline the interactions among competitors;
*Payoff*: for producer $g$ is the producer surplus $S_g^G$;
*Strategy*: for producer $g$ is the way he chooses the offers that may bring the maximal payoff in the market clearing;
*Move*: for producer $g$ is the solution of the payoff maximization problem taking into account the market clearing with the strategies of other producers fixed;
*Nash Equilibrium*:
a situation in which no producer can improve his surplus by changing his strategy while the strategies of other producers are fixed.

Let $s_g$ be the strategy of producer $g$, $\mathcal{G}_{\bar{g}}$ be the set of the producers except $g$ ($g \cup \mathcal{G}_{\bar{g}} = \mathcal{G}$), $s_{\mathcal{G}_{\bar{g}}} = \{s_i, \forall i \in \mathcal{G}_{\bar{g}}\}$: the strategy set of the competitors, $\prod_g (s_g, s_{\mathcal{G}_{\bar{g}}})$ the payoff of $g$ given the decisions of the competitors. Then, $\{s_g^*, \forall g\}$ is Nash Equilibrium if:

$$\prod_g (s_g^*, s_{\mathcal{G}_{\bar{g}}}^*) \geq \prod_g (s_g, s_{\mathcal{G}_{\bar{g}}}^*) \quad \forall g \in \mathcal{G} \tag{14}$$

In general, equilibrium can be attained by multi moves(iteration search algorithm,[4][10][14]) of the game model in which each producer solve his surplus maximization problem alternatively until no producer can improve his/her surplus by changing his strategy, given that the strategies of other producers are fixed.

### A. Supply function equilibrium (SFE) [4] [10-13]

In the SFE game models, each producer will find a linear optimal supply function to submit to the market to maximize the individual producer surplus. According to the parameterization techniques for the decision variables, three kinds of supply function models are used popularly in literature, which are listed as follows:

– *SFE-intercept*: the decision variable is $a_g$ while the $b_g$ is fixed as $b_g^m$ ($s_g = a_g$)
The supply function can be expressed as:

$$o_g(p_g) = a_g + b_g^m p_g \quad \forall g \in \mathcal{G} \tag{15}$$

– *SFE-slope*: the decision variable is $b_g$ while the $a_g$ is fixed as $a_g^m$ ($s_g = b_g$)
The supply function can be expressed as:

$$o_g(p_g) = a_g^m + b_g p_g \quad \forall g \in \mathcal{G} \tag{16}$$

– *SFE-k* parameter: the decision variable is $k_g$ and servers as a multiplier of the marginal cost ($s_g = k_g$)
The supply function can be expressed as:

$$o_g(p_g) = k_g (a_g^m + b_g^m p_g) \quad \forall g \in \mathcal{G} \tag{17}$$

Suppose the strategic supply functions take the intercept parameterization model. By applying the *KKT* conditions to the optimization problem (4) ~ (7), we can get the price of the reference bus $N$ as:

$$v_N = \frac{I_{\mathcal{G}}^T (B^m)^{-1} [J_{\mathcal{G}}^T (\mu^+ - \mu^-) + (\omega^+ - \omega^-) + a]}{I_{\mathcal{G}}^T (B^m)^{-1} I_{\mathcal{G}} - I_{\mathcal{D}}^T H^{-1} I_{\mathcal{D}}} - \frac{I_{\mathcal{D}}^T H^{-1} [J_{\mathcal{D}}^T (\mu^+ - \mu^-) + e]}{I_{\mathcal{G}}^T (B^m)^{-1} I_{\mathcal{G}} - I_{\mathcal{D}}^T H^{-1} I_{\mathcal{D}}} \tag{18}$$

The nodal prices at the generator and load buses are:

$$\lambda_{\mathcal{G}} = v_N I_{\mathcal{G}} - J_{\mathcal{G}}^T (\mu^+ - \mu^-) \tag{19}$$

$$\lambda_{\mathcal{D}} = v_N I_{\mathcal{D}} - J_{\mathcal{D}}^T (\mu^+ - \mu^-) \tag{20}$$

The subscript of the *J* matrix is to denote the corresponding rows of the *J* matrix (reduced *J* matrix). For example, $J_{\mathcal{G}}$ and $J_{\mathcal{D}}$ denote the rows of the *J* matrix corresponding to the generator buses and load buses, respectively.

The power production and the load demanded quantities are:

$$p = (B^m)^{-1} [v_N I_{\mathcal{G}} - J_{\mathcal{G}}^T (\mu^+ - \mu^-) - (\omega^+ - \omega^-) - a] \tag{21}$$

$$q = H^{-1} [v_N I_{\mathcal{D}} - J_{\mathcal{D}}^T (\mu^+ - \mu^-) - e] \tag{22}$$

With the nodal price and quantity, the maximization of the producer surplus can be formulated as:

$$\max \quad S_g^G \quad \forall g \in \mathcal{G} \tag{23}$$

$$-T \leq J(p-q) \leq T \tag{24}$$

$$P^{min} \leq p \leq P^{max} \tag{25}$$

$$<\mu^+, J(p-q) - T> = 0 \tag{26}$$

$<\mu^-, J(p-q) + T> = 0$ (27)

$<\omega^+, p - P^{max}> = 0$ (28)

$<\omega^-, -p + P^{min}> = 0$ (29)

$\mu \geq 0, \omega \geq 0;$ (30)

where the symbol "< >" denotes the element by element production of the two related vectors.

### B. Quantity bidding equilibrium

This kind of game models includes the *Cournot* [14-18] and *Stackelberg* models [19-20]. The producers will find the optimal quantity to submit to the market.

*Stackelberg* model considers leader producers who own large shares of the system capacity and are able to influence the market prices while the followers do not but can observe the quantity chosen by the leaders and select their optimal biddings. This model can be defined by a backward induction in which the leader producer would offer his quantities first and the followers take that as given. The response of the followers can be anticipated by the leaders and on that basis the leaders would decide the quantity offered.

In this paper, we discuss the *Cournot* model.

The *Cournot* model is used to analyze oligopoly markets in which the number of firms is small, or the marginal cost curve is 'steep' with respect to the demand and the size of the firms are relative similar. The decision variable is the quantity offered by each producer ($s_g = p_g$). For the maximization problem of producer $g$, the power quantities offered by other producers are assumed as given values. The nodal price at the generator bus $g$ is:

$\lambda_g = v_N - J_g^T(\mu^+ - \mu^-)$ $\forall g \in \mathcal{G}$ (31)

and $v_N = \dfrac{p_g + \sum_{i \in \mathcal{G}_g} P_i' + I_\Phi^T H^{-1}[J_\Phi^T(\mu^+ - \mu^-) + e]}{I_\Phi^T H^{-1} I_\Phi}$ (32)

where the $P_i'$ ($\forall i \in \mathcal{G}_g$) is the biding quantity of the competitors that are considered as given values derived from the last moves of corresponding producers.

The optimization problem can be expressed as:

$max \quad S_g^G$ $\forall g \in \mathcal{G}$ (33)

$-T \leq J(p - q) \leq T$ (34)

$P_g^{min} \leq p_g \leq P_g^{max}$ (35)

$<\mu^+, J(p-q) - T> = 0$ (36)

$<\mu^-, J(p-q) + T> = 0$ (37)

### C. Price bidding equilibrium

This kind of game models includes the *Bertrand* [21] and *Forchheimer* [22]. The two models respectively correspond to the *Cournot* and *Stackelberg* models; the only difference is that in the former two models the producers compete for the price while in the latter two the producers compete for the quantity. As we discussed before, under no network constraints, market clearing price is determined by the aggregate demands. Thus, since the producer will not accept negative surpluses, the price bids game among the producers will compel the producer's bid down to the marginal cost otherwise it will be substituted by other competitors who can provide lower prices, given the condition of the unlimited capacity of the producers (That is supported by the assumption that any firm can capture the entire market by pricing below others and can expand output to meet such demand [9]). If we consider the network constraints and the capacity limits, the price bidding game models are impossible to be formulated in a mathematical way since prices are actually the byproducts of the market clearing, (8), and can not be determined by the producers ex ante.

Therefore, at least for the short-term such as the hourly dispatch game with the consideration of the network constraints, the price bidding models are not suitable for the electricity markets from the analytical point of view.

### IV NUMERICAL STUDIES

As a matter of fact, the solution of the Nash equilibrium in terms of the producers' strategy is not easy due to the fact of that the sub-problem of the maximization producer surplus is a nonlinear optimization problem.

First, for the *Cournot* model, since the production of other players are fixed values, the optimization problem of the considered player is solved by sweeping all the possible states of the lines (3 lines, $3^3 = 27$ states), which makes the non-linear constraints, the complimentary equality constraints of the line flow limits (expressions (36) and (37)), transformed into linear constraints due to the fact of that the line flow states are pre-specified. The complimentary term means that, for each line $l$, either the langrage multiplier $\mu_l^+$ and $\mu_l^-$ are equal to zero with the line flow not binding or the $\mu_l^+$ ($\mu_l^- = 0$) is a positive value with the line flow binding at its limit, positive direction, or the $\mu_l^-$ ($\mu_l^+ = 0$) is a positive value with the line flow binding at its limit, negative direction.

Second, for the *SFE-intercept* model, the sweeping of the line flow states is not efficient since the production of other players are not determined (only the supply functions of the competitors are assumed as fixed in the move of the considered player). The complementary equality constraints in terms of the production limits (6 players, the production may be binding at the upper limit or lower limit or not binding, expressions (28) and (29)) and lines flows limits make the possible states of the model solution equal to $3^6 * 3^3$, which is a too large number to be solved by sweeping all the states space. In this respect, for the move of the considered player, we first find a good start point by using the heuristic optimization approach and from that start point we use the analytical approach to find the local optima around that point.

Since the nodal prices may be different when the line flows are binding, the weighted average price is introduced to represent the market clearing price. The market clearing price under constrained network can be

expressed as:

$$\underline{\lambda} = (\Sigma_g p_g \lambda_g + \Sigma_d q_d \lambda_d) / (\Sigma_g p_g + \Sigma_d q_d) \quad (38)$$

We want to point out the impacts of the physical network constraints on the market performances under strategic biddings of the producers through the market inefficiency index, the Lerner index and the allocation of surpluses among market participants.

Use the superscript $E$ and $P$ to denote the market results at the oligopoly equilibrium and perfect competition equilibrium, respectively. Use the subscript $u$ to denote the market results under unconstrained network. For example:

- $S_u^{SE}/S_u^{SP}$: Social surplus at oligopoly equilibrium/perfect competition equilibrium, without network constraints;
- $S^{SE}/\underline{\lambda}^E$: Social surplus/market clearing price at oligopoly equilibrium, with network constraints;
- $\underline{\lambda}_u^E / \underline{\lambda}_u^P$: market clearing price at oligopoly equilibrium/perfect competition equilibrium, without network constraints

The market inefficiency indices can be expressed as:

$$\xi = 100*(S^{SE} - S_u^{SP})/S_u^{SP} \quad (39)$$

$$\xi_u = 100*(S_u^{SE} - S_u^{SP})/S_u^{SP} \quad (40)$$

The Lerner indices are:

$$\sigma = (\underline{\lambda}^E - \underline{\lambda}_u^P)/\underline{\lambda}^E \quad (41)$$

$$\sigma_u = (\underline{\lambda}_u^E - \underline{\lambda}_u^P)/\underline{\lambda}_u^E \quad (42)$$

The IEEE30 bus test system is composed with 6 producers (at the 6 generator buses) and 20 consumers (at the 20 load buses), Fig. 2. The lines selected to consider the network constraints are shown in table I, other lines are assumed to have infinitive line flow limits. The parameters of the generators and the load demand curves are illustrated with the table II and III.

Perfect competition and monopoly represent the two extreme market structures, the market clearing results are illustrated with table IV and V. While in other models we assume that each generator is owned by one owner, in the monopoly case, the six generators are assumed as owned by one firm aiming at maximizing its total producer surplus, deteriorating the market performance notably with very high values of $\sigma$ (0.94), $\sigma_u$ (0.89), $\xi$ (-21.4%) and $\xi_u$ (-20%).

A more common case is the oligopoly of which the equilibrium is in-between the two preceding cases. The *Cournot* and the *SFE-intercept* game models are selected to show the oligopoly market performances under constrained and unconstrained network, the market clearing results at the oligopoly equilibrium are shown in table VI and VII. The *Cournot* model has higher values of the Lerner index and higher values of the inefficiency index (absolute value) than the *SFE-intercept* model does, both under constrained and unconstrained network, Fig. 3 and 4, suggesting the *Cournot* model possesses higher noncompetitive level than the *SFE-intercept* does. On the other hand, under the given model, with higher Lerner and inefficiency (absolute value) indices values, the constrained network brings higher level of market power than the unconstrained network does.

Fig. 5 is the producer surplus for *Cournot* model. The amount of extra surplus due to the network constraints (the total producer surplus under constrained network minus the total producer surplus under unconstrained network, 3928$–3395 $=533 $) goes to the producer $G_{22}$, $G_{23}$ and $G_{27}$, especially the producer $G_{23}$ takes the larger part. For *SFE-intercept* model, Fig. 6, only producer $G_1$ gets fewer surpluses under constrained network. Furthermore, *Counot* model contributes to higher total producer surplus, and also higher individual producer surplus, than *SFE-intercept* model does both under constrained network, 3928$ and 2725$ respectively, and under unconstrained network, 3395$ and 1835$ respectively.

Unfortunately, the favorable impacts of network constraints on the supply side are along with the adverse impacts on the consumer side. The total consumer surplus is decreased from unconstrained network case to constrained network case, Fig.7, the decrement levels are respectively 14% and 18.2% under *Cournot* and *SFE-intercept* game models.

Furthermore, under constrained network, although the market inefficiency indices of *Cournot* model and SFE-intercept model are almost the same, -6.8% and -6.4% respectively, it cannot say the two models have the same oligopoly level. Indeed, the main effects of the market power behaviors from the supply side are more remarkable under *Cournot* model with higher market clearing price, 43.8$/MW (38.5$/MW under *SFE-intercept* model), and lower exchanged power quantities, 226MW (246MW under *SFE-intercept* model). The less total producer surplus ($\Sigma_g S_g^G$) with the more total consumer surplus ($\Sigma_d S_d^C$) and mechanize surplus ($S^M$) under *SFE-intercept* game model than under *Cournot* model, table VII, results in the two models close values of the social surplus($S^S$) and thus the close values of the inefficiency indices.

Another point is that, under constrained network, the social surplus at the perfect competition equilibrium (10990 $, the last row of the table V) is even smaller than the social surplus value at the *Cournot* equilibrium under unconstrained network (11173$, the row 2 of the table VI). Therefore, to strengthen the electricity network letting it not to be congested is an imperative task that the market regulator should monitor continually, from the market efficiency point of view.

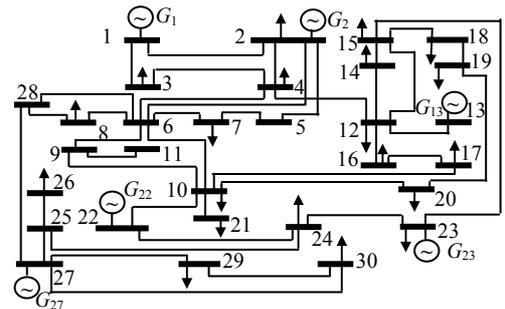

Fig.2. The IEEE30-bus transmission network

Table.I
THE CONSIDERED LINES FOR THE CONSTRAINED NETWORK

| Lines $l$ | From bus | To bus | Flow limits $T_l$ MW |
|---|---|---|---|
| 7 | 4 | 6 | 5 |
| 25 | 10 | 20 | 5 |
| 33 | 24 | 25 | 5 |

Table II
PARAMETERS FOR THE PRODUCERS

| Bus $g$ | $a_g^m$ \$/MW | $b_g^m$ \$/MW$^2$ | $P_g^{min}$ MW | $P_g^{max}$ MW |
|---|---|---|---|---|
| 1 | 25 | 0.15 | 5 | 80 |
| 2 | 20 | 0.25 | 5 | 60 |
| 13 | 23 | 0.2 | 5 | 60 |
| 22 | 22 | 0.25 | 5 | 60 |
| 23 | 20 | 0.2 | 5 | 80 |
| 27 | 22 | 0.15 | 5 | 70 |

Table III
PARAMETERS FOR THE LOAD DEMAND CURVES

| Bus $d$ | $e_d$ \$/MW | $f_d$ \$/MW$^2$ | bus $d$ | $e_d$ \$/MW | $f_d$ \$/MW$^2$ |
|---|---|---|---|---|---|
| 2 | 125 | -5 | 17 | 100 | -4.5 |
| 3 | 80 | -4 | 18 | 80 | -4 |
| 4 | 100 | -4 | 19 | 100 | -5 |
| 7 | 150 | -5 | 20 | 100 | -5 |
| 8 | 120 | -4.5 | 21 | 75 | -3.5 |
| 10 | 100 | -4 | 23 | 70 | -3 |
| 12 | 120 | -5 | 24 | 80 | -4.5 |
| 14 | 80 | -3.5 | 26 | 80 | -4 |
| 15 | 80 | -3 | 29 | 75 | -4 |
| 16 | 80 | -4 | 30 | 100 | -5 |

Table.IV
THE MARKET EQUILIBRIUM UNDER MONOPOLY (MONO.) AND PERFECT COMPETITION (PERF.), UNCONSTRAINED NETWORK

| | $S^S$ \$ | $\lambda$ \$/MW | $\sum_g P_g$ MW | $\sum_g S_g^G$ \$ | $\sum_d S_d^C$ \$ | $S^M$ \$ |
|---|---|---|---|---|---|---|
| Mono. | 9120 | 59.8 | 158 | 5601 | 3519 | 0 |
| Perf. | 11351 | 31.6 | 295 | 1439 | 9912 | 0 |

Table.V
THE MARKET EQUILIBRIUM UNDER MONOPOLY (MONO.) AND PERFECT COMPETITION (PERF.), CONSTRAINED NETWORK

| | $S^S$ \$ | $\lambda$ \$/MW | $\sum_g P_g$ MW | $\sum_g S_g^G$ \$ | $\sum_d S_d^C$ \$ | $S^M$ \$ |
|---|---|---|---|---|---|---|
| Mono. | 8924 | 61.3 | 152 | 5506 | 3321 | 97 |
| Perf. | 10990 | 32.6 | 280 | 1465 | 9247 | 279 |

Table VI
THE MARKET EQUILIBRIUM UNDER COURNOT (COUT.) AND SFE-INTERCEPT (SFE) MODELS, UNCONSTRAINED NETWORK

| | $S^S$ \$ | $\lambda$ \$/MW | $\sum_g P_g$ MW | $\sum_g S_g^G$ \$ | $\sum_d S_d^C$ \$ | $S^M$ \$ |
|---|---|---|---|---|---|---|
| Cout. | 11173 | 39.4 | 258 | 3395 | 7778 | 0 |
| SFE | 11345 | 33 | 288 | 1835 | 9509 | 0 |

Table VII
THE MARKET EQUILIBRIUM UNDER COURNOT (COUT.) AND SFE-INTERCEPT (SFE) MODELS, CONSTRAINED NETWORK

| | $S^S$ \$ | $\lambda$ \$/MW | $\sum_g P_g$ MW | $\sum_g S_g^G$ \$ | $\sum_d S_d^C$ \$ | $S^M$ \$ |
|---|---|---|---|---|---|---|
| Cout. | 10581 | 43.8 | 226 | 3928 | 6380 | 273 |
| SFE | 10618 | 38.5 | 246 | 2725 | 7416 | 477 |

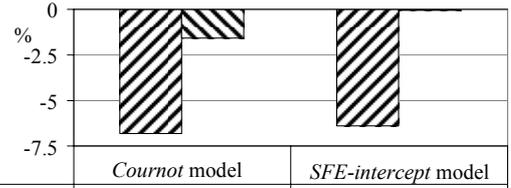

| | *Cournot* model | *SFE-intercept* model |
|---|---|---|
| Constrained network $\zeta$ (%) | -6.8 | -6.4 |
| Unconstrained network $\zeta_u$ (%) | -1.57 | -0.06 |

Fig.3. The market inefficiency indices

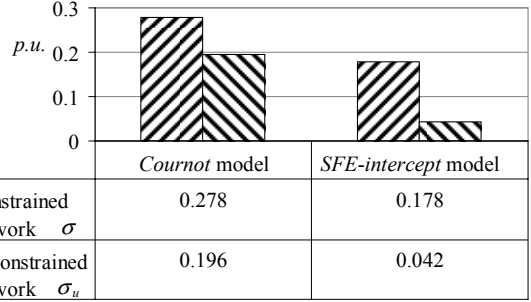

| | *Cournot* model | *SFE-intercept* model |
|---|---|---|
| Constrained network $\sigma$ | 0.278 | 0.178 |
| Unconstrained network $\sigma_u$ | 0.196 | 0.042 |

Fig.4. The Lerner indices

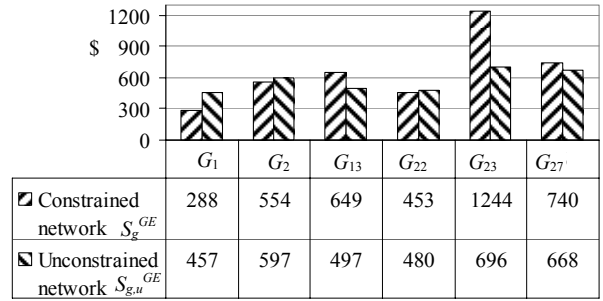

| | $G_1$ | $G_2$ | $G_{13}$ | $G_{22}$ | $G_{23}$ | $G_{27}$ |
|---|---|---|---|---|---|---|
| Constrained network $S_g^{GE}$ | 288 | 554 | 649 | 453 | 1244 | 740 |
| Unconstrained network $S_{g,u}^{GE}$ | 457 | 597 | 497 | 480 | 696 | 668 |

Fig.5. The surplus of different producers under *Cournot* model

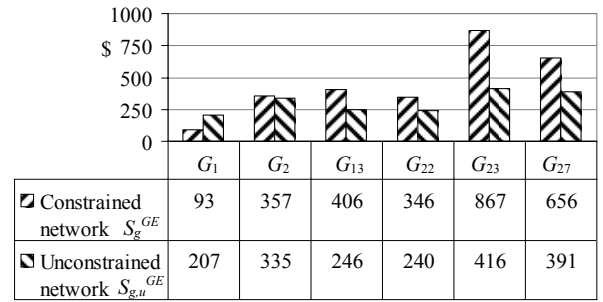

| | $G_1$ | $G_2$ | $G_{13}$ | $G_{22}$ | $G_{23}$ | $G_{27}$ |
|---|---|---|---|---|---|---|
| Constrained network $S_g^{GE}$ | 93 | 357 | 406 | 346 | 867 | 656 |
| Unconstrained network $S_{g,u}^{GE}$ | 207 | 335 | 246 | 240 | 416 | 391 |

Fig.6. The surplus of different producers under *SFE-intercept* model

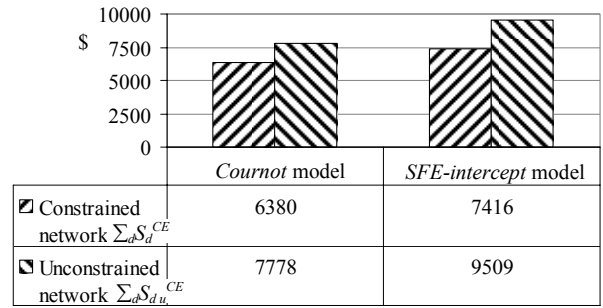

| | *Cournot* model | *SFE-intercept* model |
|---|---|---|
| Constrained network $\sum_d S_d^{CE}$ | 6380 | 7416 |
| Unconstrained network $\sum_d S_{d,u}^{CE}$ | 7778 | 9509 |

Fig.7. The consumer surplus under *Cournot* and *SFE-intercept* models

## V. CONCLUSIONS

For the specific features of the electricity industry, the present electricity markets may be better described in terms of oligopoly than of perfect competition from which they may be rather far. Game theory is an appropriate tool to model electricity markets in an oligopoly competitive environment.

In the electricity markets, in which the power transactions are undertaken on a grid that needs to be operated under strict physical and operational constraints; for this reason very specific occasions of market power behaviors related to system congestion may arise, giving a further source of market inefficiency.

Game theory models suitable to represent the competitive electricity market have been analyzed and tested using the IEEE30 bus system. The simulations show a worsening of the market performance, as measured by the inefficiency index and the Lerner index, when compared to the ideal model of perfect competition. Effects generated by oligopoly competition are: loss in total social surplus, increase of the producer surplus at the expenses of consumer surplus, decrease in the quantity exchanged on the market, higher market clearing price.

The loss of efficiency with respect to the perfect competition may vary a lot from a model to another. This shows that the type of competition in the market and/or the hypothesis we do to model it may lead to very different results. The *monopoly* model shows the worst behavior, both under constrained and unconstrained network. As for the oligopoly market, *Counort* model show a worse performance than the *SFE-intercept* model, and as a fact that, from the simulation results of other oligopoly models, *Cournot* model has the least competitive level both under constrained and unconstrained network.

Due to the network constraints, the transmission network plays a major role to determine the market equilibrium. Under constrained network, the market clearing price is higher and the power transacted is lower than the corresponding values under unconstrained network. As for the surpluses values, the network constraints provide some producers with additional opportunity to get higher surplus with the decrement of the consumer surplus, leading to the higher level of market inefficiency compared with the unconstrained network. In this respect, to strengthen the electricity transmission network will contribute to mitigate the market power behaviors of the electricity producers.